\documentclass[a4paper,english,aps,preprint]{revtex4}
\usepackage[T1]{fontenc}
\usepackage[latin9]{inputenc}
\usepackage{babel}

\usepackage[unicode=true, pdfusetitle,
 bookmarks=true,bookmarksnumbered=false,bookmarksopen=false,
 breaklinks=false,pdfborder={0 0 1},backref=false,colorlinks=false]
 {hyperref}
\hypersetup{
 citebordercolor={0 1 0}}
 
\makeatletter
\@ifundefined{textcolor}{}
{%
 \definecolor{BLACK}{gray}{0}
 \definecolor{WHITE}{gray}{1}
 \definecolor{RED}{rgb}{1,0,0}
 \definecolor{GREEN}{rgb}{0,1,0}
 \definecolor{BLUE}{rgb}{0,0,1}
 \definecolor{CYAN}{cmyk}{1,0,0,0}
 \definecolor{MAGENTA}{cmyk}{0,1,0,0}
 \definecolor{YELLOW}{cmyk}{0,0,1,0}
 }

\makeatother

\makeatother

\makeatother

\makeatother

\makeatother

\makeatother

\makeatother

\makeatother

\makeatother

\makeatother

\makeatother

\makeatother

\usepackage{babel}

\makeatother

\usepackage{babel}

\makeatother

\usepackage{babel}

\makeatother

\usepackage{babel}

\makeatother

\usepackage{babel}

\makeatother

\usepackage{babel}

\makeatother

\begin{document}

\title{Relativistic corrections to the Sellmeier equation allow derivation
of Demjanov's formula}

\author{Peter C. Morris}

\email{ptr.mrrs@optusnet.com.au}

\affiliation{Transputer Systems, PO Box 544, Marden, Australia 5070}
\begin{abstract}
A recent paper by V.V. Demjanov in Physics Letters A reported a formula
that relates the magnitude of Michelson interferometer fringe shifts
to refractive index and absolute velocity. We show that relativistic
corrections to the Sellmeier equation allow an alternative derivation
of the formula. 
\end{abstract}
\maketitle

\section{Introduction}

In Sellmeier dispersion theory\cite{Sellmeier}, a material is treated
as a collection of atoms whose negative electron clouds are displaced
from the positive nucleus by the oscillating electric fields of the
light beam. The oscillating dipoles resonate at a specific frequency
(or wavelength), so the dielectric response is modeled as one or more
Lorentz oscillators. For $m$ oscillators the Sellmeier equation may
be written.

\begin{equation}
n^{2}=1+\sum_{i=1}^{m}\frac{A_{i}\lambda^{2}}{\lambda^{2}-\lambda_{i}^{2}}\label{eq:SellGeneral}\end{equation}

Here $\lambda$ is the wavelength of the incident light in vacuum.
But if the vacuum is physical and our frame of reference undergoes
length contraction due to motion relative to that vacuum, then $\lambda$
for that direction will be proportionally shorter than in our frame
of reference. To correct for that effect, $\lambda$ in the above
formula might need to replaced with $\lambda/\gamma$, where $\gamma$
is the Lorentz gamma factor.

This possibility is explored below. To simplify calculations we will
assume that one oscillator dominates the contribution to $n$. We
can then use the following one term form of the Sellmeier equation,\begin{equation}
n^{2}=1+\frac{B\lambda^{2}}{\lambda^{2}-C}\label{eq:SellOneTerm}\end{equation}

where, if oscillator $j$ is the dominant one,

\[
C=\lambda_{j}^{2}\]

and

\begin{equation}
B=B_{j}=\sum_{i=1}^{m}\;\frac{\lambda^{2}-\lambda_{j}^{2}}{\lambda^{2}-\lambda_{i}^{2}}\; A_{i}\label{eq:B}\end{equation}

accounts for contributions to $n$ from all the $A_{i}$, at the specific
wavelength of $\lambda$.

We will then find that replacing $\lambda$ by $\lambda/\gamma$ and
making one further change, allows derivation of Demjanov's formula\cite{Demjanov,DemjanovMMX,Dmitriyev}.

\section{Derivation}

\subsection{Replacing $\lambda$ with $\lambda/\gamma$}

Let a Michelson interferometer containing an optical medium be moving
with absolute velocity $v$ in the direction of its parallel arm.

We hypothesize that the refractive index for a light path through
the medium is given by the following formula, derived from (\ref{eq:SellOneTerm})
above,

\[
n=\sqrt{1+\frac{B\left(\lambda/\gamma_{p}\right)^{2}}{\left(\lambda/\gamma_{p}\right)^{2}-C}}\]

\begin{equation}
=\sqrt{1+\frac{B\lambda^{2}}{\lambda^{2}-C\gamma_{p}^{2}}}\label{eq:1}\end{equation}

where 
\begin{itemize}
\item $\gamma_{p}=\frac{1}{\sqrt{1-\frac{v_{p}^{2}}{c^{2}}}}$ and, 
\item $v_{p}$ is the component of absolute velocity that is parallel to
the light path 
\end{itemize}
Solving for $\lambda$ gives the inverse relation,

\begin{equation}
\lambda=\sqrt{\frac{C\gamma_{p}}{1-\frac{B}{n^{2}-1}}}\label{eq:2}\end{equation}

Let the refractive index for the orthogonal arm for light of wavelength
$\lambda_{0}$ be $n$.

Then since $v_{p}$ for the orthogonal arm is zero, equation (\ref{eq:2})
gives,

\begin{equation}
\lambda_{0}=\sqrt{\frac{C}{1-\frac{B}{n^{2}-1}}}\label{eq:3}\end{equation}

Then using equation (\ref{eq:1}), the refractive index for the parallel
arm, moving with absolute velocity $v_{p}$ must be,

\[
n_{p}=\sqrt{1+\frac{B\lambda_{0}^{2}}{\lambda_{0}^{2}-C\gamma_{p}^{2}}}\]

\begin{equation}
=\sqrt{1+\frac{B}{1-\frac{C}{\lambda_{0}^{2}}\gamma_{p}^{2}}}\label{eq:4}\end{equation}

Substituting $v_{p}=v$ and also for $\lambda_{0}$ and $\gamma_{p}$
gives,

\begin{equation}
n_{p}=\sqrt{1+\frac{B\left(1-\frac{v^{2}}{c^{2}}\right)}{\frac{B}{(n^{2}-1)}-\frac{v^{2}}{c^{2}}}}\label{eq:5}\end{equation}

For $v\ll c$ Taylor expansion then gives,

\begin{equation}
n_{p}=n-\frac{(n^{2}-1)(B+1-n^{2})}{2Bn}\:\frac{v^{2}}{c^{2}}+O(v^{4})\label{eq:6}\end{equation}

As $n$ and $n_{p}$ are effective refractive indexes in the frame
of the interferometer, we can now calculate travel times for return
light paths as follows.

For the orthogonal arm,

\[
T_{o}=\frac{2L}{c/n}=\frac{2Ln}{c}\]

For the parallel arm,

\[
T_{p}=\frac{2L}{c/n_{p}}=\frac{2Ln_{p}}{c}\]

giving a travel time difference of

\[
\Delta T=T_{o}-T_{p}=\frac{2L}{c}\:(n-n_{p})\]

\begin{equation}
=\frac{L}{c}\:\frac{(n^{2}-1)(B+1-n^{2})}{Bn}\:\frac{v^{2}}{c^{2}}+O(v^{4})\label{eq:8}\end{equation}

which using $\epsilon=n^{2}$ and $\Delta\epsilon=n^{2}-1$ can be
re-expressed as,

\begin{equation}
\Delta T=\frac{v^{2}}{c^{2}}\;\frac{L}{c\sqrt{\epsilon}}\;\frac{\Delta\epsilon(B-\Delta\epsilon)}{B}\label{eq:DeltaT}\end{equation}

When $B\approx1$ this approximates Demjanov's formula (9) in \cite{Demjanov},
which is,

\begin{equation}
\Delta T=\frac{v^{2}}{c^{2}}\;\frac{L}{c\sqrt{\epsilon}}\;\Delta\epsilon(1-\Delta\epsilon)\label{eq:Demjanov}\end{equation}

\subsection{Derivation of a closer variant}

The above steps obtained a variant of Demjanov's formula. However
substituting the Sellmeier coefficients for water provided in Table
1 of \cite{Coello} into equations (\ref{eq:SellGeneral}) and (\ref{eq:B}),
showed that the value of $B$ would be less than the value of $n^{2}-1=\Delta\epsilon$
for wavelengths of light%
\footnote{Eg. For light of wavelength $900\; nm$ through water, $B=0.755745$
while $\Delta\epsilon=0.760544$%
} used by Demjanov with light paths through water. So formula (\ref{eq:DeltaT})
predicts negative values of $\Delta T$ for water, while Demjanov's
formula predicts positive values. The experimentally determined values
shown in Figure 1 of \cite{Demjanov} are positive, so for the case
of water, formula (\ref{eq:DeltaT}) is falsified.

This prompted an effort to obtain a closer variant and version 5 of
this paper describes the obtaining of,

\begin{equation}
\Delta T=\frac{v^{2}}{c^{2}}\;\frac{L}{c\sqrt{\epsilon}}\;\frac{\Delta\epsilon(1-\Delta\epsilon)}{B}\label{eq:Variant}\end{equation}

However, even this is not close enough because for gases\cite{Griesmann},
B is orders of magnitude smaller than 1, so predictions for $\Delta T$
obtained using (\ref{eq:Variant}) would be much larger than observed
values\cite{Demjanov,DemjanovMMX,Miller}.

So another attempt was made, but as the next section shows, this resulted
in derivation of Demjanov's formula itself.

\subsection{Derivation of Demjanov's formula}

This section reports how Demjanov's formula was obtained by considering
hypothetical effects of absolute motion on the orthogonal arm, in
addition to those assumed for the parallel arm.

For non-polarized light, the dipoles within the material would oscillate
in a plane orthogonal to the light beam and the direction of absolute
motion would also lie in that plane. The amplitude and frequency of
oscillations parallel with the direction of absolute motion might
be affected and those effects might distribute to oscillations in
other directions within the plane as well.

Lacking knowledge of such effects, a simple approach was to experiment
with applying different kinds of correction until a satisfactory formula
for $\Delta T$ was obtained. Eventually the following was found,

\begin{equation}
n=\sqrt{1+\frac{Bg_{o}^{2}\lambda^{2}}{\lambda^{2}-C\gamma_{p}^{2}g_{o}^{2}}}\label{eq:B1a}\end{equation}

where 
\begin{itemize}
\item $g_{o}=1+\left(1-B\right)(\gamma_{o}-1)$ is hypothesized to account
for effects of orthogonal absolute motion, 
\item $\gamma_{o}=\frac{1}{\sqrt{1-\frac{v_{o}^{2}}{c^{2}}}}$ , and, 
\item $v_{o}$ is the component of absolute velocity that is orthogonal
to the light path. 
\end{itemize}
The above definition of $g_{o}$ has the following features 
\begin{itemize}
\item When $v_{o}=0$, $g_{o}=1$ 
\item For $v_{o}\ll c$ , $g_{o}^{2}=1+\left(1-B\right)\frac{v_{o}^{2}}{c^{2}}+O(v^{4})$
\end{itemize}
Since the quantity $Bg_{o}^{2}$ is related to oscillation amplitude,
this is equivalent to a hypothesis that absolute motion causes a negative
feedback effect when $B>1$ and a positive feedback effect when $B<1$.

\subsubsection*{Continuing the derivation}

Let the refractive index for the medium at rest in the physical vacuum
for light of wavelength $\lambda_{0}$ be $n$.

Then since $v_{o}$ and $v_{p}$ are zero, solving (\ref{eq:B1a})
for $\lambda$ gives,

\begin{equation}
\lambda_{0}=\sqrt{\frac{C}{1-\frac{B}{n^{2}-1}}}\label{eq:B3}\end{equation}

Then for the orthogonal arm, substituting $\lambda=\lambda_{0}$ and
$\gamma_{p}=1$ (since $v_{p}=0$) in equation (\ref{eq:B1a}), gives
the refractive index as,

\[
n_{o}=\sqrt{1+\frac{Bg_{o}^{2}}{1-\left(1-\frac{B}{n^{2}-1}\right)g_{o}^{2}}}\]

For $v_{o}=v\ll c$, we can substitute $g_{o}^{2}\approx1+\left(1-B\right)\frac{v^{2}}{c^{2}}$
which gives,

\[
n_{o}=\sqrt{1+\frac{B\left(1+(1-B)\frac{v^{2}}{c^{2}}\right)}{1-\left(1-\frac{B}{n^{2}-1}\right)\left(1+(1-B)\frac{v^{2}}{c^{2}}\right)}}\]

and Taylor expansion then gives,

\begin{equation}
n_{o}=n-\frac{(n^{2}-1)^{2}(B-1)}{2Bn}\:\frac{v^{2}}{c^{2}}+O(v^{4})\label{eq:B5o}\end{equation}

For the parallel arm using $v_{o}=0$ and $v_{p}=v$ we once again
obtain equation (\ref{eq:6}) viz.,

\begin{equation}
n_{p}=n-\frac{(n^{2}-1)(B+1-n^{2})}{2Bn}\:\frac{v^{2}}{c^{2}}+O(v^{4})\label{eq:B6}\end{equation}

Using the same logic as before,

\[
T_{o}=\frac{2L}{c/n_{o}}=\frac{2Ln_{o}}{c}\]

\[
T_{p}=\frac{2L}{c/n_{p}}=\frac{2Ln_{p}}{c}\]

giving travel time difference,

\[
\Delta T=T_{o}-T_{p}=\frac{2L}{c}\:(n_{o}-n_{p})\]

\[
=\frac{2L}{c}\:\left(\left(n-\frac{(n^{2}-1)^{2}(B-1)}{2Bn}\:\frac{v^{2}}{c^{2}}\right)-\left(n-\frac{(n^{2}-1)(B+1-n^{2})}{2Bn}\:\frac{v^{2}}{c^{2}}\right)\right)+O(v^{4})\]

\[
=\frac{L}{c}\:\left(\frac{(n^{2}-1)(-Bn^{2}+B+n^{2}-1)}{Bn}+\frac{(n^{2}-1)(B+1-n^{2})}{Bn}\right)\:\frac{v^{2}}{c^{2}}\]

\begin{equation}
=\frac{L}{c}\:\frac{(n^{2}-1)(2-n^{2})}{n}\:\frac{v^{2}}{c^{2}}\label{eq:B8}\end{equation}

which using $\epsilon=n^{2}$ and $\Delta\epsilon=n^{2}-1$ is Demjanov's
formula,

\begin{equation}
\Delta T=\frac{v^{2}}{c^{2}}\;\frac{L}{c\sqrt{\epsilon}}\;\Delta\epsilon(1-\Delta\epsilon)\label{eq:BDeltaT}\end{equation}

\section{Conclusion}

The above has shown that relativistic corrections to the Sellmeier
equation allow derivation of Demjanov's formula.

However it should be noted that the corrections presented here are
not the only possible ones. So it remains to be seen if they are valid
or whether better alternatives will be found.
\begin{acknowledgments}
The author is grateful to Dr. V.P. Dmitriyev for helpful comments. \end{acknowledgments}

\end{document}